\documentclass{article}
\usepackage{epsfig}

\begin{document}
\baselineskip .7cm

\author{Navin Khaneja \thanks{To whom correspondence may be addressed. Email:navin@hrl.harvard.edu}\thanks{Division of Engineering and Applied Sciences, Harvard University, Cambridge, MA 02138} \ \ and \ 
Niels Chr. Nielsen \thanks{To whom correspondence may be addressed. Email: ncn@inano.dk}\thanks{Center for Insoluble Protein Structures (inSPIN), Interdisciplinary Nanoscience Center (iNANO) and Department of Chemistry, University of Aarhus, DK-8000 Aarhus C, Denmark}}

\vskip 4em

\title{\bf Triple oscillating field technique for accurate measurements of couplings in homonuclear spin systems}

\maketitle


\begin{center} {\bf Abstract} \end{center}

We present a new concept for homonuclear dipolar recoupling in magic-angle-spinning (MAS) solid-state NMR experiments which avoids the problem of dipolar truncation. This is accomplished through the introduction of a new NMR pulse sequence design principle: the triple oscillating field technique. We demonstrate this technique as an efficient means to accomplish broadband dipolar recoupling of homonuclear spins, while decoupling heteronuclear dipolar couplings and anisotropic chemicals shifts and retaining influence from isotropic chemical shifts. In this manner, it possible to synthesize Ising interaction ($2I_zS_z$) Hamiltonians in homonuclear spin networks and thereby avoid dipolar truncation - a serious problem essentially all previous homonuclear dipolar recoupling experiments suffer from. Combination of this recoupling concept with rotor assisted dipolar refocusing enables easy readout of internuclear distances through comparison with analytical Fresnel curves. This forms the basis for a new class of solid-state NMR experiments with potential for structure analysis of  uniformly $^{13}$C labelled proteins through accurate measurement of $^{13}$C-$^{13}$C internuclear distances. The concept is demonstrated experimentally by measurement of C$_\alpha$-C', C$_\beta$-C', and C$_\gamma$-C' internuclear distances in powder samples of the amino acids \textit{L}-alanine and \textit{L}-threonine.

\vskip 0.5cm

{\bf Abbreviations:} TOFU, triple oscillating field technique; RADAR, rotor assisted dipolar refocusing; MAS, magic angle spinning; CP, cross polarization. 

\vskip 0.5cm

\section{Introduction}

For more than a decade, dipolar truncation has been considered a major obstacle for the use of solid-state NMR spectroscopy to determine structures of biological macromolecules \cite{griffinnatstruct,hodgkinsonemsley,kiihne,oschkinatnature,hohwyband,smith,seashore,levitttrunc}. Dipolar truncation occurs when a spin, say $I_1$, is strongly coupled to a spin $I_2$ (usually a nearby spin of the same species) and weakly coupled to another spin $I_3$ (a distant spin) and the two coupling Hamiltonians do not commute. The result is that the strong coupling between $I_1$-$I_2$  truncates (averages) the coupling between $I_1$-$I_3$, with the consequence that the long-range dipolar interaction cannot be measured directly. This problem is particularly serious for the measurement of internuclear distances in homonuclear spin systems of uniformly isotope-labelled molecules (e.g., $^{13}$C-$^{13}$C couplings), since essentially all dipolar recoupling experiments presented to date (e.g., \cite{rotres,drama,rfdr,horror,c7,baba,postc7,spc5,levittencycl,dream}) generate double- or zero-quantum dipolar Hamiltonians, which cause non-commuting interactions in multiple spin systems. This has shifted the focus of accurate distance measurements in uniformly $^{13}$C,$^{15}$N-labelled proteins to heteronuclear dipolar recoupling experiments \cite{rienstradenovo,jaroniecfibril}, although they often suffer from smaller dipole-dipole couplings due to coupling of $^{13}$C to nuclei such as $^{15}$N with a lower gyromagnetic ratio. In basic heteronuclear dipolar recoupling experiments such as REDOR \cite{redor}, TEDOR \cite{tedor}, and the $\gamma$-encoded variant GATE \cite{gate}, dipolar truncation is avoided since the large chemical shift difference between hetero spins allows for selective manipulation of one of the heteronuclei making it possible to prepare an Ising Hamiltonian $2I_zS_z$ which can be maintained in dipolar recoupling experiments. This Hamiltonian commutes across different spins pairs and therefore does not suffer from dipolar truncation.  

The importance of the homonuclear dipolar truncation problem can be seen in the vast majority of studies addressing structure determination of isotope-labelled biomolecules by solid-state NMR. Either the long-range couplings are missing due to truncation or the long-range connections are assessed indirectly through the involvement of spin diffusion via heteronuclear spins \cite{protassistdiff,darr,fujiwara,mcdermott,baldus,ernst,pain}. To circumvent this problem, many strategies such as spin system dilution by special isotope labelling procedures \cite{oschkinatnature,hong} or the use of arrays of experiments which selectively or band-selectively recouple only specific dipolar coupling interactions \cite{hohwyband,seashore,r2trterao,r2trcosta,r2w,meierr2,r2trw,tyckor2} may be used to establish long-range structural constraints. In all cases, there is a price to pay. This may be lack of precise distance constraints, disturbing influence from relaxation parameters and multiple-spin effects, need for replicating the experiments many times for different isotope labelling patterns or using experiments with different setting of "selectivity-inducing" pulse sequence parameters. No solution has so far been presented, which in a broadband manner allows recoupling of a $2I_zS_z$ type of Hamiltonian for homonuclear spin systems in proteins in presence of highly abundant heteronuclear spins such as $^1$H. A recent paper by Levitt and coworkers \cite{levitttrunc} suggested recoupling of the native dipolar coupling Hamiltonian simultaneously with chemical shielding interactions to truncate the transverse parts of the dipolar Hamiltonian. This solution to the dipolar truncation problem also recouples heteronuclear dipolar couplings and thereby imposes difficulties in practical implementation for samples that are not fully deuterated.

In this paper, we introduce the {\bf T}riple {\bf O}scillating {\bf F}ield techniq{\bf U}e (TOFU) which allows for broadband recoupling of Ising interaction Hamiltonian, $2I_zS_z$, while simultaneously ensuring that the chemical shifts of the spins are maintained and the heteronuclear couplings are decoupled. We show how this technique circumvents the dipolar truncation problem and forms the basis for accurate distance measurement by solid-state NMR. The paper is organized as follows. In the first section, we develop the TOFU principle for design of homonuclear dipolar recoupling rf pulse sequence elements. In the subsequent section, we demonstrate the power of the concept by developing experiments which use TOFU recoupling along with  {\bf R}otor {\bf A}ssisted {\bf D}ipol{\bf A}r {\bf R}efocusing  (RADAR) to establish long-range two-spin dipolar coupling interactions in multiple-spin systems. The exclusive selection of two-spin dipolar interactions and elimination of relaxation effects by combination of results from two experiments provides an easy way to read out the internuclear distances graphically by comparing signal intensities with a map of Fresnel curves representing different internuclear distances. These features are demonstrated numerically and  experimentally by measuring internuclear distances in uniformly $^{13}$C,$^{15}$N-labelled samples of \textit{L}-alanine and \textit{L}-threonine serving as models for multiple homonuclear spin systems.

\section{Homonuclear Recoupling without Dipolar Truncation by Triple Oscillating Fields}

Consider two coupled homonuclear spins $I$ and $S$ under magic angle spinning (MAS) conditions. In a Zeeman frame, rotating with the spins at their Larmor frequency, the Hamiltonian of the spin system takes the form 
\begin{equation}
\label{Eq:rotH0} 
H_0(t) = \omega_{I}(t) I_z +  \omega_{S}(t) S_z +  \omega_{IS}(t) (3I_zS_z - \overline{I} \cdot \overline{S}) \quad ,
\end{equation}
where $\omega_{I}(t)$ and $\omega_{S}(t)$ represents the time-varying chemical shifts for the two  spins $I$ and $S$ respectively, and $\omega_{IS}(t)$ the coupling between them. These interaction strengths may be expressed in terms of a Fourier series 
\begin{equation}
\label{eq:fourier}
\omega_{\lambda}(t) = \sum_{m=-2}^{2}\omega_{\lambda}^{(m)}\exp(im\omega_r t) \quad ,
\end{equation}
where $\omega_r$ is the sample spinning frequency (in angular units), while the coefficients $\omega_{\lambda}^{(m)}$ $(\lambda = I, S, IS)$ reflect dependences on physical parameters like the isotropic chemical shift, the anisotropic chemical shift, the dipole-dipole coupling constant and through this the internuclear distance \cite{simpson}. 

\begin{figure}[h]
\begin{center}
\includegraphics[scale=.7]{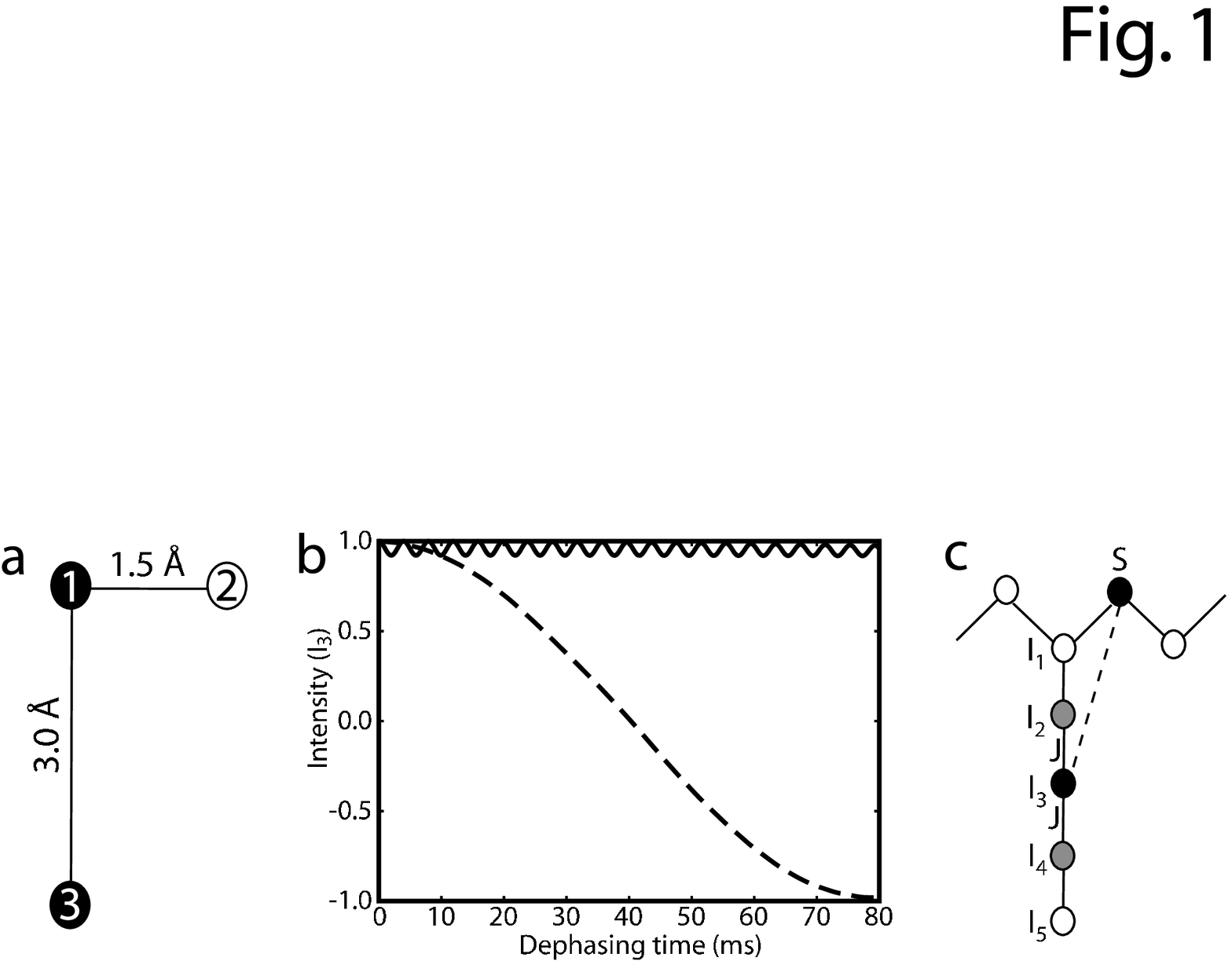}
\end{center}
\caption{Illustration of the effect of dipolar truncation. (a) Three-spin topology with spin 1 strongly coupled to spin 2 and weakly coupled to a distant spin 3.  (b) Dephasing curves for the magnetization on spin 3 when subject to conventional planar recoupling (solid-line, POST-C7 pulse sequence \cite{postc7}) and to Ising type recoupling (dashed line, TOFU pulse sequence element (\textit{vide infra})). (c) Typical spin topology consisting of an S spin (in the present work $^{13}$C') along a number of I spins (in the present work $^{13}$C$_\alpha$, $^{13}$C$_\beta$ etc.) encountered in measurement of internuclear distances in uniformly $^{13}$C-labelled proteins. The curves in (b) are calculated using the spin-topology in (a) for a single-crystal with orientation (0,45,0), dipolar couplings of size ($b_{IS}/2\pi$)/orientation ($\alpha_{PC},\beta{PC},\gamma{PC}$ in degrees) of I$_1$-I$_2$: -2100 Hz/(0,0,0), I$_1$-I$_3$: -300 Hz/(0,90,0), and isotropic shifts relative to the rf carrier of -15.5 kHz (I$_3$), 0 Hz (I$_2$), and -12 kHz (I$_1$). The spinning frequency was 20 kHz, and the experimental parameters as described in the caption to Fig. 3.}  
\label{fig:1}
\end{figure}

In absence of radio-frequency (rf) irradiation, fast MAS averages the dipolar part of the coupling $\omega_{IS}(t)$  to zero. This averaging may be prevented using dipolar recoupling techniques, among which the conventional schemes exploit the fact that $\omega_{IS}(t)$ contains modulations at frequencies $\pm \omega_r$ and $\pm 2 \omega_r$. Such terms may be demodulated by irradiating the spin system with an rf field whose amplitude is set to, e.g., $\frac{1}{2}\omega_r$ and the phase to say $x$ \cite{horror}. This rotates the coupling tensor in the interaction frame of the rf irradiation at frequency $\omega_r$ and thereby demodulates the $\omega_r$ modulation arising from the rotor revolution.  This forms the basis of many homonuclear recoupling sequences \cite{horror,c7,postc7,spc5,levittencycl,dream}, which in this manner prepare an effective  \textit{planar} recoupling Hamiltonian of the kind $ \bar{H} = D (2I_zS_z - 2I_yS_y)$ where $D$ is a constant depending on the  dipole-dipole coupling and a scaling induced through recoupling. In a network of coupled spins, such planar Hamiltonians between spin pairs do not commute. For transfer of coherence between directly coupled spins (e.g., for spectral assignment purposes) this is acceptable, but when it comes to measurement of the structurally important long-range interactions this non-commutative feature is fatal. The recoupled planar Hamiltonian between the distant spins  $I_1$ and $I_3$ is truncated by a stronger planar Hamiltonian between  $I_1$ and  $I_2$ implying that the most important information about the spin system is effectively lost. This feature is illustrated in Fig. \ref{fig:1}, showing a typical three-spin topology along with a dipolar dephasing curve (solid line) for conventional planar recoupling. The lack of dephasing of the spin $I_3$ signal is a signature of dipolar truncation.

With the aim of solving the dipolar truncation problem, we take inspiration from the heteronuclear recoupling experiments and demonstrate how we can recouple spins $I_1$ and $I_3$ in the presence of $I_2$, by synthesizing mutually commuting Ising ($2I_zS_z$) dipolar Hamiltonians. In this manner we can establish efficient dipolar dephasing through the long-range dipolar coupling as illustrated by the dashed line in Fig. \ref{fig:1}b. The underlying principle is to use the chemical shift difference between spins to truncate the recoupled Hamiltonian to an Ising  Hamiltonian. This is accomplished by demodulation of $\omega_{IS}(t)$ in such a manner that the isotropic chemical shifts of the spins is preserved $and$ the heteronuclear dipolar couplings are eliminated. This requires extension of existing recoupling strategies with additional modulations, which here for the first time is demonstrated by introducing the Triple Oscillating Field technique, TOFU.

\begin{figure}[h]
\begin{center}
\includegraphics[scale=.6]{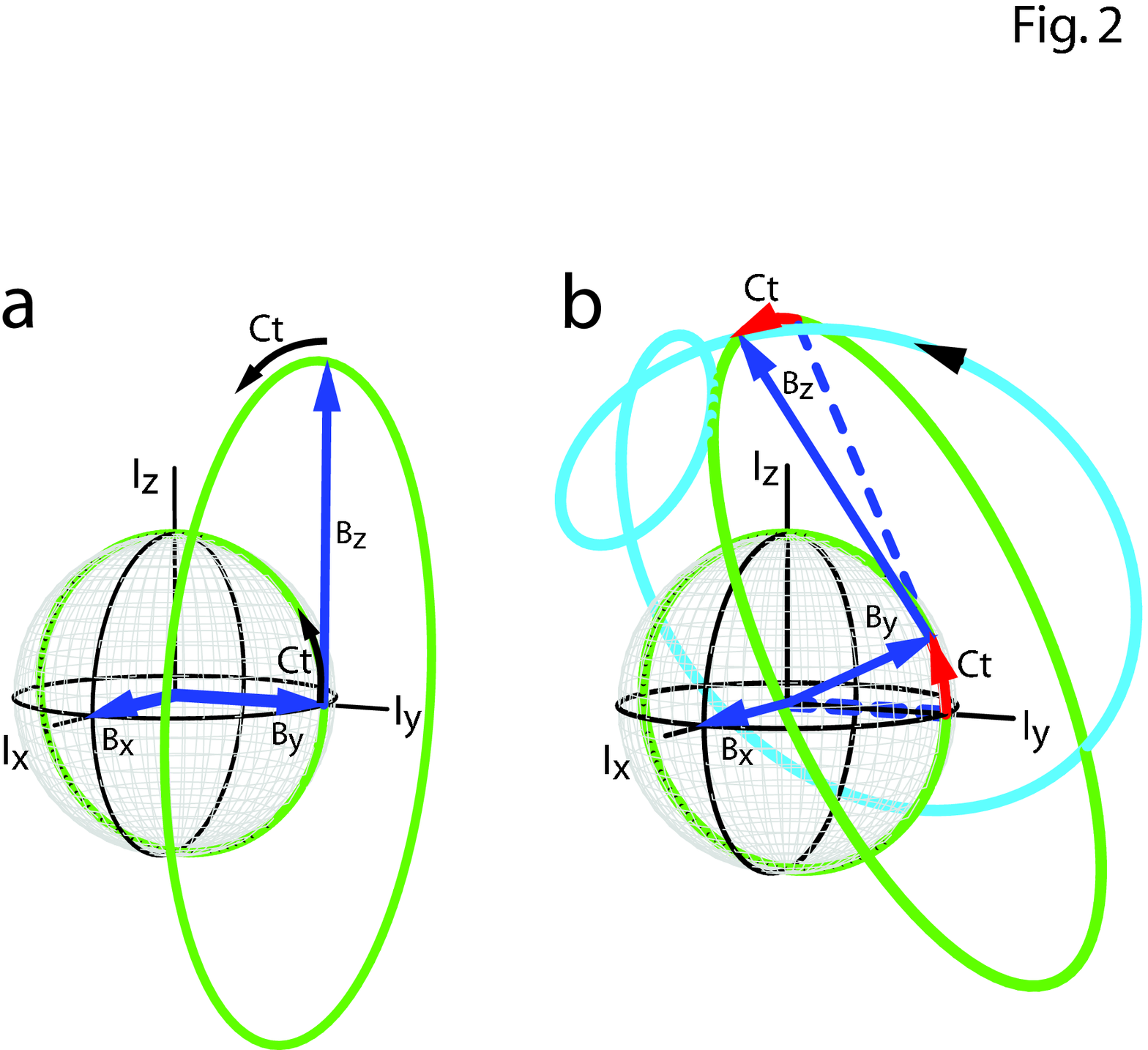}
\end{center}
\caption{Graphical representation of the triple oscillating field for the TOFU dipolar recoupling experiment with the three field components represented by solid blue vectors. In the Zeeman frame, there is a constant field $B_x = C$ along the $x$ axis, a field $B_y(t)$ of strength $C$ that rotates around $B_x$ at angular frequency $C$, and a field $B_z(t)$ of strength $B$ that rotates around $B_y(t)$ at angular frequency $C$. At time $t=0$, $B_y$ points in the $y$ direction and $B_z$ in the $z$ direction as illustrated in (a). The green curve shows the instantaneous trajectories for the $B_y$ and $B_z$ field components. (b) shows the situation at a time corresponding to $Ct = \pi/6$ (the red arrows indicates the directions for the movement of the field vectors from time 0). In a more accurate representation, the light blue curve shows the true movement of the tip of the $B_z$ vector affected by the simultaneous $Ct$ rotation around the two axes. We note that in practice $B$ (the length of the vector $B_z$) is much larger than $C$ (the length of the vectors $B_x$ and $B_y$).}
\label{fig:2}
\end{figure}

The spin system is irradiated by an rf field characterized by an amplitude $A(t)$, a phase $\phi(t)$, and an offset $\omega(t)$. In Zeeman frame, the rf Hamiltonian takes the form
\begin{equation}
H_{rf}(t)  = A(t) (\cos \phi(t)  F_x + \sin \phi(t) F_y) + \omega(t) F_z \quad ,
\end{equation}
where $F_q=I_q+S_q$ ($q=x,y,z$). The amplitude, the phase, and the offset are chosen in such a way that the resulting field looks like a constant field $B_x$ of strength $C$ along the $x$-axis, a field $B_y$ of strength $C$ that rotates around $B_x$ with angular frequency $C$, and a field $B_z$ of strength $B$ rotating around $B_y$ with angular frequency $C$ as depicted in Fig. \ref{fig:2}. The resulting rf field Hamiltonian can be written as
\begin{eqnarray}
\label{eq:phase.design}
\nonumber H_{rf}(t) &=& C F_x + C \exp(-i C t F_x ) \ F_y \ \exp(i C t F_x) \\ 
 &+&   B  \exp(-i C t F_x)\exp(-i C t F_y) \ F_z \ \exp(i C t F_y)\exp(i C t F_x) \quad . 
\end{eqnarray}
 This particular setting of the rf fields may be appreciated by performing a series of co-ordinate transformations. In the interaction frame of the $x$-phase rf irradiation (we call this the demodulation frame), the rf Hamiltonian transforms to $ C F_y + B \exp(-iC t F_y) F_z \exp(iC t F_y)$. Moving now into the interaction frame of the $y$-phase irradiation with strength $C$ (we call this the recoupling frame), the rf Hamiltonian transforms to $ B F_z $. The overall transformation to the recoupling frame may be represented as $\tilde{A} = U^+(t) A U(t)$ with $U(t)=\exp(- i C t F_x) \exp(- i C t F_y)$. Within this frame, we can establish expressions for the chemical shielding and dipolar coupling interactions to find appropriate conditions for truncation-free dipolar recoupling.

When transformed into the recoupling frame, the \textit{chemical shielding} Hamiltonian takes the form
\begin{equation}
\label{eq:trans}
\tilde{H}_\sigma(t) = \omega_{I}(t) ( \cos^2(Ct) I_z + \sin(Ct) I_y - \frac{1}{2} \sin(2Ct) I_x )
+ \omega_{S}(t) ( \cos^2(Ct) S_z  + \sin(Ct) S_y - \frac{1}{2} \sin(2Ct) S_x ) 
\end{equation}
and the \textit{dipolar coupling} Hamiltonian takes the form
\begin{eqnarray}
\label{eq:mainprem}
\nonumber \tilde{H}_{IS}(t) &=& \frac {3}{2} \omega_{IS}(t) \{  \cos^4(Ct) 2I_zS_z  +  
\sin^2(Ct)\cos^2(Ct) 2I_xS_x + \sin^2(Ct) 2 I_yS_y \\ 
\nonumber  &+&   \frac{1}{2}\sin(2Ct) [\cos(Ct) (2I_zS_y + 2I_yS_z) -\sin(Ct) (2I_xS_y + 2I_yS_x) \\ &-& \cos^2(Ct) (2I_zS_x + 2I_xS_z) ] \}  \quad ,
\end{eqnarray}
where we ignored components from the invariant (and non-recoupled) $\overline{I} \cdot \overline{S}$ part of the Hamiltonian in Eq. (\ref{Eq:rotH0}). This part is not rotated by the rf field and averages out under MAS.

By recalling that the rf Hamiltonian in the recoupling frame is $ B F_z $, if $B$ is chosen sufficiently large (i.e., $B \gg C$ and $B \gg \omega_r$), then the parts of the Hamiltonians that do not commute with $F_z$ is averaged out. For efficient truncation of transverse terms in  Eq. (\ref{eq:trans}), by the rf-field $BF_z$, we should have $|B \pm 2C| \gg |\omega_I^{(0)}|$, $|B \pm (2C+\omega_r)| \gg |\omega_I^{(\pm 1)}|$, and $|B \pm 2(C+\omega_r)| \gg |\omega_I^{(\pm 2)}|$. The field $B$, when chosen to satisfy these constraints is also sufficiently strong to average parts of the dipolar interactions that do not commute with $BF_z$, as their strengths typically are weaker than the chemical shifts. The averaged Hamiltonian, in the interaction frame of $BF_z$, takes the form,
\begin{eqnarray}
\label{eq:main}
\nonumber  H_A (t) &=& \omega_{I}(t)  \cos^2(Ct) I_z + \omega_{S}(t)  \cos^2(Ct) S_z \\ 
&+& \frac{3}{2} \omega_{IS}(t) \{ \cos^4(Ct) (3I_zS_z - \overline{I} \cdot \overline{S}) +  (I_xS_x + I_yS_y) \} \quad .
\end{eqnarray}
We note that in the recoupling frame, the effective rf Hamiltonian $B F_z$ and the scaled isotropic chemical shifts both contributes to the averaging in Eq. (\ref{eq:main}), implying that the most efficient truncation will depend on the rf carrier position. Interference between time dependence (of the kind $\exp(-im \omega_r t)$) of the interactions frequencies (c.f., Eq. (\ref{eq:fourier})) and the time modulation $\cos^2(Ct)=\frac{1}{2} + \frac{1}{4} \exp(i2Ct)+ \frac{1}{4} \exp(-i2Ct)$ and $\cos^4(Ct)=\frac{3}{8} + \frac{1}{4} \exp(i2Ct)+ \frac{1}{4} \exp(-i2Ct)+ \frac{1}{16} \exp(i4Ct)+ \frac{1}{16} \exp(-i4Ct)$, arising from the pulse sequence, allows for selective demodulation (i.e., recoupling) of specific terms in the Hamiltonian in Eq. (\ref{eq:main}) by fulfilling the conditions $nC+m\omega_r$=0 (where  $m$ = $\pm 1, \pm 2$ according to Eq. (\ref{eq:fourier}) and $n$ = $\pm 2, \pm 4$).

We will here consider two classes of recoupling experiments, obtained using \textbf{(a)} $C = \frac{1}{4}\omega_r$, and \textbf{(b)} $C = \frac{1}{2}\omega_r$. These two experiments are qualitatively different in terms of the interactions they recouple, thereby not only illustrating the flexibility of the TOFU principle but also  introducing two different methods applicable for samples with and without dominant heteronuclear coupling interactions to abundant spins (\textit{vide infra}).

If we choose $C = \frac{1}{4}\omega_r$, then the dipolar term $\cos^4(Ct)$ has oscillating components at frequencies $\pm \frac{1}{2}\omega_r$ and $\pm \omega_r$, while the chemical shift term $\cos^2(Ct)$ only has oscillating components at frequencies $\pm \frac{1}{2}\omega_r$. The oscillating part at $\pm \omega_r$, generates stationary (secular) terms with $\omega_{IS}(t)$ under MAS conditions, while no stationary terms is generated for the chemical shift interaction. When retaining only the secular components in Eq. (\ref{eq:main}), we get 
\begin{eqnarray}
\label{eq:main1}
\overline{{H_A}}= \frac{1}{2} \omega_{I}^{(0)} I_z + \frac{1}{2} \omega_{S}^{(0)} S_z  + 
\frac{3}{16}c_{IS}^{(1)} \cos \gamma \{ 2I_zS_z - I_xS_x - I_yS_y \} \quad.
\end{eqnarray}
We note that under MAS conditions, $\omega_I^{(0)}$ and $\omega_S^{(0)}$ represents the isotropic chemical shifts of the spins $I$ and $S$ and $\omega_{IS}^{(1)}+\omega_{IS}^{(-1)} = 2 c_{IS}^{(1)} \cos(\gamma)$, where $\gamma$ is the phase of rotation of internuclear vector around the 
rotor axis. The dipolar Fourier factor $c_{IS}^{(1)}$ is related to the dipole-dipole coupling constant $b_{IS} = -\frac{\mu_0}{4\pi}\frac{\gamma_I \gamma_S}{r^3_{IS}}$ (and thereby to the internuclear distance $r^3_{IS}$ and the gyromagnetic ratios $\gamma_I$ and $\gamma_S$) as $c_{IS}^{(1)} = \frac{1}{2 \sqrt{2}} b_{IS} \sin(2 \beta)$, where $\beta$ is the Euler angle between the inter-nuclear vector and the rotor axis \cite{simpson}.

The planar terms $I_xS_x + I_yS_y$ of the Hamiltonian are usually a source of dipolar truncation. But note, the effective Hamiltonian in Eq. (\ref{eq:main1}) retains the isotropic chemical shift of nuclei and therefore in presence of large magnetic fields, the difference in the chemical shifts $\omega_I^{(0)} - \omega_S^{(0)}$ will in most cases of practical relevance truncate the planar terms $I_xS_x + I_yS_y$. On the other hand, the effective Hamiltonian does not include components from anisotropic chemical shielding, nor heteronuclear dipolar couplings. This implies that the effective Hamiltonian in Eq. (\ref{eq:main1}) can be used to recouple homonuclear spins while avoiding the problem of dipolar truncation in uniformly $^{13}$C-labelled samples with $^1$H and $^{15}$N spins present in high abundance. 

If we choose $C = \frac{1}{2}\omega_r$, then the dipolar term $\cos^4(Ct)$ has oscillating components at frequencies $\pm \omega_r$ and $\pm 2 \omega_r$, both of which will generate secular terms under MAS conditions. Similarly, the chemical shift term $\cos^2(Ct)$ has oscillating components at $\pm \omega_r$, which will recouple the anisotropic chemical shielding and heteronuclear dipolar coupling interactions under MAS conditions. Overall the recoupling experiment will lead to a mixture of two different types of homonuclear dipolar components and as described by the dipolar recoupled Hamiltonian
\begin{eqnarray}
\label{eq:main1Chalf}
\nonumber \overline{H_A}&=& \frac{1}{2} (\omega_{I}^{(0)} + \frac{1}{2} (\omega_I^{(1)}+\omega_I^{(-1)}))I_z + \frac{1}{2}(\omega_{S}^{(0)}  + \frac{1}{2} (\omega_S^{(1)}+\omega_S^{(-1)})) S_z  \\
&+&
 \frac{3 }{4} ( c_{IS}^{(1)} \cos \gamma +\frac{1}{4} c_{IS}^{(2)} \cos 2 \gamma ) \{ 2I_zS_z - I_xS_x - I_yS_y \} \quad ,
\end{eqnarray}
where the upper line represents chemical shielding terms and the lower dipolar coupling terms. For the latter, we note that $\omega_{IS}^{(2)}+\omega_{IS}^{(-2)} = 2 c_{IS}^{(2)} \cos(2\gamma)$ with $c_{IS}^{(2)} = -\frac{1}{4} b_{IS} \sin^2(\beta)$. The planar terms of the recoupled dipolar Hamiltonian in practice will be truncated by the chemical shift terms. We observe that ($i$)  the dipolar scaling factor of the first term is four times larger than for the $C = \frac{1}{4}\omega_r$ experiment, ($ii$) the period of the TOFU element is shorter by a factor of 2, but also that ($iii$) this experiment recouples anisotropic shielding and more problematic heteronuclear dipolar couplings. These are generally undesirable and need to be removed for measurement of distances between the spins unless deuterated samples are used. In this respect, the experiment ${\bf (b)}$ resemble the experiment proposed by Levitt and coworkers \cite{levitttrunc} and may prove  useful in the sense that the anisotropic shielding may assist  dipolar truncation and the dipolar factor is sufficiently large that relatively long $^{13}$C-$^{13}$C distances may be probed. In the subsequent analysis, we will only treat experiment \textbf{(a)} as it does not recouple heteronuclear dipolar interactions being problematic for the most relevant biological samples with $^1$H in natural abundance.

\begin{figure}[h]
\begin{center}
\includegraphics[scale=.6]{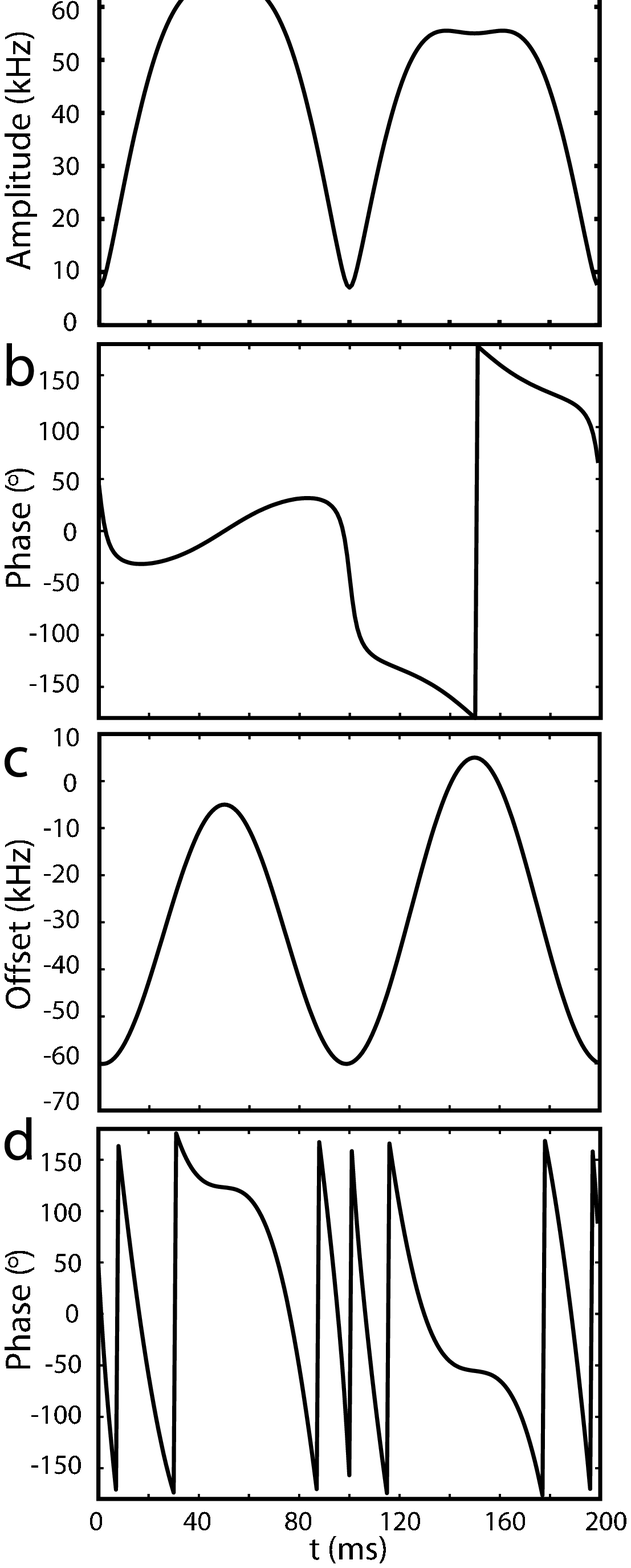}
\end{center}
\caption{Schematic illustration of a TOFU pulse sequence element for the case of $C=\frac{1}{4}\omega_r$ as represented by Eq.(\ref{eq:phase.design}). (a) The amplitude $A(t)$, (b) the phase $\phi(t)$, (c) the instantaneous offset $\omega(t)/2\pi$, and (d) the total phase $\phi(t) + \int_0^{t} \omega(\tau) d\tau$ in the Zeeman frame which implements the desired offset $-\omega(t)$ for applications with constant carrier frequency. The duration of the element is four rotor periods, and the curves were generated for the case of $\omega_r/\pi$ = 20 kHz and $B=3\omega_r$.}
\label{fig:3}
\end{figure}

As a representative example of a TOFU dipolar recoupling element, Figure \ref{fig:3}, shows the amplitude, phase, and offset of the rf field for the specific case of $C = \frac{1}{4}\omega_r$, $B = 3 \omega_r$, and $\omega_r/2\pi$ = 20 kHz.

\subsection{RADAR: Rotor Assisted Dipolar Refocusing}

\begin{figure}[h]
\begin{center}
\includegraphics[scale=.7]{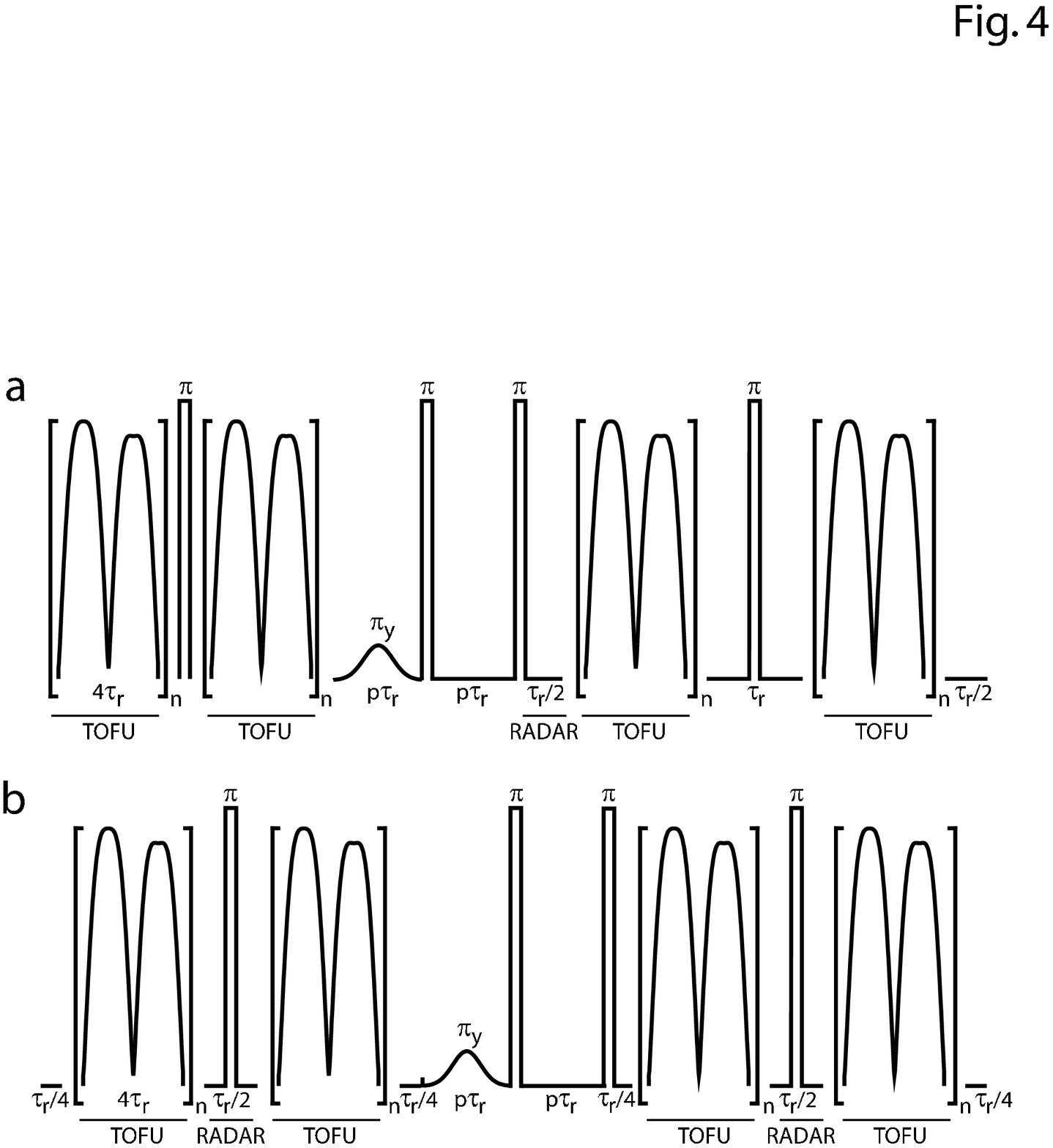}
\end{center}
\caption{Pulse sequences combining TOFU and RADAR elements for measurement of homonuclear internuclear distances in biological solids (see typical spin topology in Fig. \ref{fig:1}c). (a) The main experiment and (b) the reference experiment (see text). Dipolar dephasing curves are measured by incrementing all TOFU building block in steps of $4 \tau_r$ ($2 \tau_r$ for the experiment with $C=\frac{1}{2}\omega_r$ expressed in terms of the overall time variable $T=n 16 \tau_r$. All rf pulses are of phase $x$ unless denoted otherwise. $p$ is an integer chosen to rotor synchronize the selective inversion element (presently chosen as a Gaussian pulses). The $\tau_r/4$ periods in the reference experiment ensures that the pulse sequence is symmetrical with respect to the chemical shift refocusing $\pi$ pulses and that the pulse sequences are of equal length. The experiments are typically preceded by a $^1$H to $^{13}$C cross-polarization pulse sequence element preparing $x$-phase coherence on all $^{13}$C spins.}
\label{fig:4}
\end{figure}

In this section, we show how the TOFU pulse sequence element may be implemented experimentally to measure distances in homonuclear spin systems of biological relevance. We focus on measuring the distances of $n$ spins $I$, from a given $S$ spin as illustrated by the spin topology in Fig. \ref{fig:1}c. The TOFU recoupling results in a Hamiltonian that takes the form
\begin{equation}
H = \sum_{jk} \pi J_{jk} 2 I_{jz}I_{kz} + \sum_{jk} \omega_{jk}^{DD} 2 I_{jz}I_{kz} + 
\sum_{k} \pi J_{k} 2 I_{kz}S_{z} + \sum_{k} \omega_{k}^{DD} 2 I_{kz}S_{z}, 
\end{equation}
where $J_{jk}$ represents the strength of scalar couplings (in Hz) of spin $I_j$ to spin $I_k$, $J_k$ the scalar coupling between spins $k$ and spin $S$, $\omega_{jk}^{DD}$ the strength of the recoupled dipole-dipole coupling between $I_j$ and $I_k$, and  $\omega_{k}^{DD}$  the recoupled dipolar couplings between spin $k$ to spin $S$. We note that both $\omega_{jk}^{DD}$ and $\omega_{k}^{DD}$ have orientation dependence reflecting the orientation of the individual crystallites in the powder sample as described in the previous section. 

We measure the dipolar coupling $\omega_{k}^{DD}$ by observing the dephasing of the transverse magnetization of spin $I_k$ under this coupling. For this purpose, it is desirable to refocus all the other dipolar couplings $\omega_{jk}^{DD}$, as these are orientation dependent and rapidly dephase the coherence on the $I$ spins, which becomes difficult to separate from dephasing caused by the coupling to spin $S$ of interest. Our strategy is to perform a reference experiment (Fig. \ref{fig:4}b), in which all dipolar couplings and the scalar couplings $J_{k}$ are refocused. The magnetization of a spin $I_k$, only evolves under the $J$ coupling to surrounding $I$ spins. Then we perform the main experiment (Fig. \ref{fig:4}a), in which $\omega_{jk}^{DD}$ and  $J_{k}$ are refocused, however, the recoupled Ising coupling $\omega_{k}^{DD}$ between the spin $I_k$ and spin $S$ is maintained. This is achieved by exploiting the fact that the recoupled dipolar couplings have a $\cos(\gamma)$ dependency (see Eq. (\ref{eq:main1})). This implies that if the Hamiltonian $\omega_{jk}^{DD} I_{jz}I_{kz}$ is allowed to evolve for time $\frac{T}{2}$, followed by delay of $\frac{\tau_r}{2}$ (i.e., half a rotor period), and then evolution in another $\frac{T}{2}$ period, then the dipolar coupling is refocused as the half rotor period delay reverses the sign of recoupled dipolar Hamiltonian. We note that a similar approach was used in the experiment proposed by Levitt and coworkers \cite{levitttrunc}. For practical reasons, we henceforth denote this method of Rotor Assisted DipolAr Refocusing as RADAR. The dipolar coupling $\omega_{k}^{DD}$ between spin $I_k$ and $S$ can be protected from being refocused by performing a selective $\pi$ pulse on spin $S$ after time $\frac{T}{2}$. The selective $\pi$ pulse on spin $S$ also refocuses the scalar couplings $J_{k}$.

The schematics for two experiments is shown in Figure \ref{fig:4}. In the reference experiment a delay of half rotor period is introduced in the center of each $\frac{T}{2}$ evolutions to refocus all recoupled dipolar Hamiltonians. Both experiments are performed by incrementing $T$ (each of the four TOFU elements are incremented in units of $4$ rotor periods for the experiment with $C=\frac{1}{4}\omega_r$ and $2$ rotor periods for the experiment with $C=\frac{1}{2}\omega_r$). We observe the dephasing of spin $I_k$ as a function of $T$ in both the experiments and extract $\omega_{k}^{DD}$ and hence the distance between spin $I_k$ and $S$ from this as described below. 

The transverse $x$-magnetization of spin $I_k$ in the reference experiment dephases as  
\begin{equation}
S_r(T) = \exp(-\lambda T) \Pi_{j=1}^n \cos(\pi J_{jk}^{DD}T ) \quad , 
\end{equation}
where $\lambda$ is the transverse relaxation rate. In the main experiment, the $x$ magnetization of spin $I_k$ dephases as 
\begin{equation}
 S_m(t) = \exp(-\lambda T) \Pi_{j=1}^n \cos(\pi J_{jk}^{DD}T ) \int \cos(\omega_{k}^{DD} T) \quad , 
\end{equation}
where the integration denotes powder averaging. Now we define 
\begin{equation}
\label{fresnel}
 \eta(T) = \frac{S_r(t)-S_m(t)}{S_r(t)} = 1 - \int \cos(\omega_{k}^{DD} T) \quad . 
\end{equation}
$\eta(T)$ is a typical Fresnel curve that depends exclusively on the desired coupling $\omega_{k}^{DD}$ in a manner previously seen in the heteronuclear case for REDOR experiments with the signal detected on the most abundant spin (i.e., $^{13}$C) \cite{redor,mueller}. The internuclear distances between spin I$_k$ and S can now be read directly by comparison of experimental data with curves calculated for different dipolar coupling constants. We should note, that in case of perfect dipolar truncation these expressions are exact and intrinsically impose several advantages relatively to previous techniques for measurement of homonuclear dipolar coupling constants in diluted spin systems or by using frequency selective pulse sequences: (\textit{i}) the dephasing is not dependent on relaxation such as the frequently encountered zero- or double-quantum relaxation, (\textit{ii}) the method is strictly reflecting dephasing under one anisotropic interaction - the dipole-dipole coupling of interest, and (\textit{iii}) it does not suffer appreciably from influence from anisotropic shielding tensor magnitudes and orientations. 

\section{Experimental}

All experiments were performed on a Bruker Avance 400 NMR spectrometer ($^1$H Larmor frequency of 400 MHz) equipped with a triple resonance 2.5 mm probe operating in double-resonance mode. The data for uniformly $^{13}$C,$^{15}$N-labelled samples of $L$-alanine and $L$-threonine (purchased from Cambridge Isotope Laboratories, Andover, MA) were obtained using the full volume of standard 2.5 mm rotors (12 $\mu$l) at ambient temperature using 20 kHz sample spinning. The experiments used 4 s relaxation delay, 4 scans, 1.5 ms $^1$H to $^{13}$C cross polarization (CP) \cite{cp} with rf field strengths of 55 ($^1$H) and 35 ($^{13}$C) kHz, and an rf field strength of 80 kHz for the non-selective $^{13}$C refocusing pulses. SPINAL-64 \cite{spinal64} decoupling with an rf field strength of 105 kHz was used during acquisition, while continuous-wave  decoupling with an rf field strength of 140 kHz was used during the TOFU-RADAR period. Selective inversion of the $^{13}$C' spins were accomplished using a Gaussian pulse of duration 250 $\mu$s. The TOFU waveform was implemented with the carrier on-resonance at the $^{13}$C' spin (offset incorporated in phase as in Fig. \ref{fig:3}d) with the rf irradiation digitized in 200 steps for each 4 $\tau_r$ period. All simulations were performed using the open-source SIMPSON \cite{simpson,simpson2,simpson3,simpson4} software, with representative $^{13}$C chemical shielding and dipolar coupling tensors established using the open-source software SIMMOL \cite{simmol}. Unless specified otherwise, all simulations used REPULSION \cite{repulsion} powder averaging with 66-144 $\alpha_{PR}$, $\beta_{PR}$ crystallite angles and 5-10 $\gamma_{PR}$ angles.

\section{Results and Discussion}

\begin{figure}[h]
\begin{center}
\includegraphics[scale=.7]{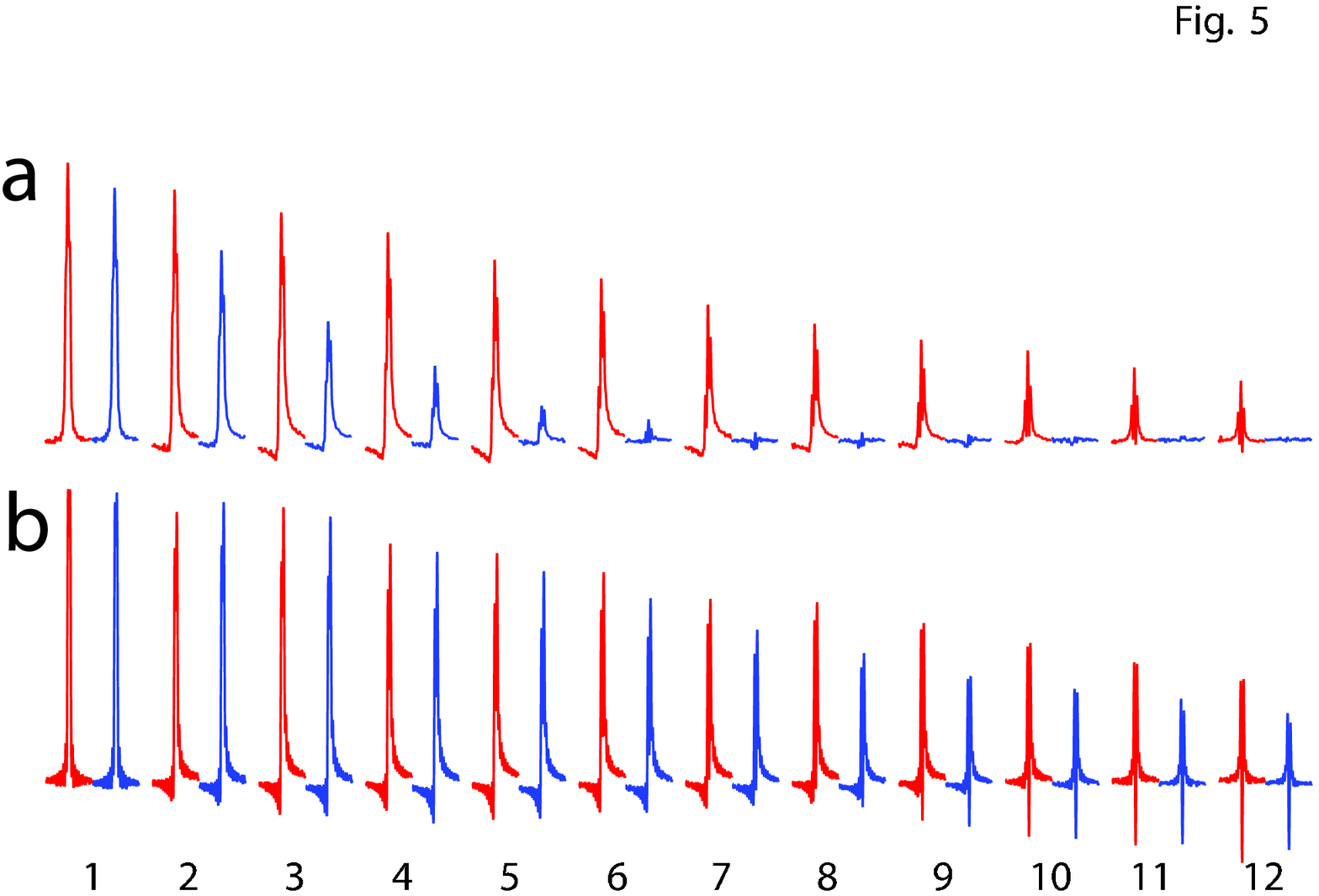}
\end{center}
\caption{Experimental TOFU-RADAR dephasing spectra obtained for the $^{13}$C$_\alpha$ (a) and $^{13}$C$_\beta$ (b) for a powder of $L$-alanine using the main experiment in Fig. \ref{fig:4}a (blue line) and the reference experiment in Fig. \ref{fig:4}b (red line) with $C=\frac{1}{4}\omega_r$, $B = 3 \omega_r$, and 20 kHz spinning and incrementation of $T$ in steps of 16 (4x4) rotor periods from 16 $\tau_r$ (left) to 240$\tau_r$ (right).}
\label{fig:5}
\end{figure}

To illustrate the capability of the TOFU-RADAR experiments for determination of internuclear distances in biomolecules, Figure \ref{fig:5}, shows a series of experimental dipolar dephasing spectra for the $^{13}$C$_\alpha$ and $^{13}$C$_\beta$ carbons of a powder of uniformly $^{13}$C,$^{15}$N-labelled $L$-alanine with the carbonyl acting as the dephasing S spin. The spectra were obtained at a 100 MHz ($^{13}$C Larmor frequency) using 20 kHz spinning and the rf fields adjusted to the $C=\frac{1}{4}\omega_r$ and $B=3\omega_r$ condition. The carrier was placed on-resonance on the $^{13}$C' resonance to provide efficient truncation (c.f., Eq. (\ref{eq:main1})). From the spectra, it is clearly evident that ($i$) the main TOFU-RADAR experiment (Fig. \ref{fig:4}a) leads to faster dephasing than the reference experiment (Fig. \ref{fig:4}b) and ($ii$) that the $^{13}$C$_\alpha$ carbon dephases much faster than the $^{13}$C$_\beta$ carbon due to a larger dipole-dipole constant (i.e., smaller distance) to the dephasing $^{13}$C' spin. We note that the spectra display minor variations in the phase of the signal upon increasing the $T$ period which, however, does not seem influence significantly the extraction of information about the dipolar coupling constant and thereby the internuclear distances upon integrating the overall signal intensities as demonstrated below.

\begin{figure}[h]
\begin{center}
\includegraphics[scale=.6]{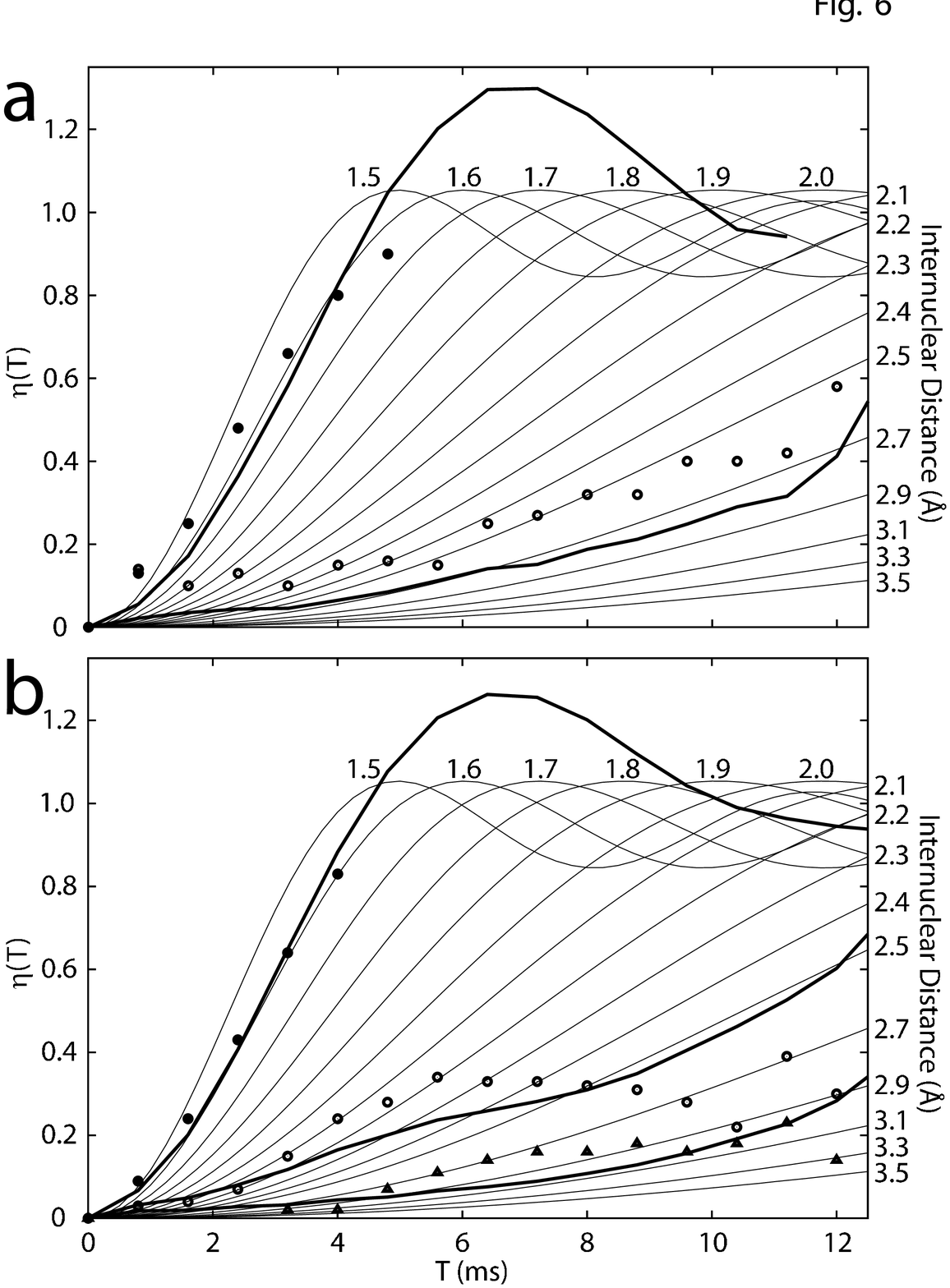}
\end{center}
\caption{Demonstration of the use of the TOFU-RADAR experiments for determination of internuclear distances. Experimental $\eta(T)$ data for $^{13}$C$_\alpha$ (filled circle), $^{13}$C$_\beta$ (open circle), and $^{13}$C$_\gamma$ (triangle) along with analytical Fresnel curves (thin solid lines) for different internuclear distances and numerically exact SIMPSON simulations (solid curves).}
\label{fig:6}
\end{figure}

The straightforward extraction of internuclear distances from TOFU-RADAR experiments linear combined according to Eq. (\ref{fresnel}) is demonstrated in Fig. \ref{fig:6} which compares experimental and simulated $\eta(T)$ curves for $L$-alanine and $L$-threonine. The $\eta(T)$ data for $^{13}$C'-$^{13}$C$_\alpha$, $^{13}$C'-$^{13}$C$_\beta$, and $^{13}$C'-$^{13}$C$_\gamma$ dephasing were obtained using the same experimental conditions as in Fig. \ref{fig:5}. The experimental $\eta(T)$ data can be used for direct extraction of distances by overlaying them on a chart of analytical Fresnel curves corresponding to different internuclear distances in Eq. (\ref{fresnel}). The reasonable match between the experimental curves and the analytical curves for nominal $^{13}$C'-$^{13}$C$_\alpha$, $^{13}$C'-$^{13}$C$_\beta$, and $^{13}$C'-$^{13}$C$_\gamma$ distances of 1.5, 2.5, and 2.9 \AA \ clearly demonstrates the effectiveness of this proposed experiment and the simple readout of distances from the Fresnel chart. For $L$-alanine the $^{13}$C'-$^{13}$C$_\alpha$ distance has been measured to 1.536 \AA \ and 1.537 \AA \ using X-ray diffraction \cite{destro} and neutron diffraction \cite{lehman}, while the $^{13}$C'-$^{13}$C$_\beta$ distance was measured to 2.525 and 2.532 \AA using the methods. Both values aggrees remarkably well (within 0.1 \AA) with the data obtained using the Fresnel-curve interpretted TOFU-RADAR curves. For $L$-threonine the $^{13}$C'-$^{13}$C$_\alpha$, $^{13}$C'-$^{13}$C$_\beta$, and $^{13}$C'-$^{13}$C$_\gamma$ distances has been determined to 1.56 (1.54), 2.47 (2.55), and 2.93 (3.09) \AA \ using rotational resonance based solid-state NMR spectroscopy \cite{meierr2} and X-ray diffraction (number in parenthesis) \cite{janczak} which also fits very well with the estimations from Fig. \ref{fig:6}c. The good aggreement clearly documents the power of dipolar truncation-free recoupling and it demonstrates it may be possible to obtain the most important distance information from recording only the main experiment and the reference experiment at a single appropriately chosen time $T$. This would enable straightforward integration of this distance measurement method with multiple-dimensional experiments for improved resonance resolution. Studies along these lines are presently in progress.

While this visual approach should be adequate for reasonably precise distance measurements in multiple dimensional experiments, we should note that the experimental data do not match perfectly with the analytical Fresnel curves. The minor deviations may be ascribed to residual effects from anisotropic shielding partly being introduced by non-perfect selective inversion of the $^{13}$C' magnetization due to the relatively large chemical shielding anisotropy. Such effects may be appreciated by comparison with numerical simulations as included in Fig. \ref{fig:6}. Indeed these curves may account for some of the "Fresnel curve transversing" and modulation behavior of the experimental data. Minor remaining deviations between the calculated curves and the experimental data may be ascribed to non-perfect estimation of all tensorial interactions in the multiple-spin systems and deviations from perfect selective inversion and $^1$H decoupling throughout the long dephasing periods. Overall, we find the match between the experimental and simulated curves sufficiently good to prove our concept for accurate distance measurements. We estimate the error bars in the distance measurement to be less than $\pm 0.2$ \AA when comparing with either simulated curves or with the map of Fresnel curves. This is a very high accuracy when comparing it to the standard tolerances typically accepted in liquid-state NMR structure determination (where distances typically are estimated to be within bands of 1.8 - 2.8, 2.8 - 3.8, and above 3.8 \AA) or solid-state NMR relying on proton or rf driven spin diffusion.

\section{Conclusion and Outlook}

In conclusion, we have demonstrated for the first time that it is possible to obtain homonuclear dipolar recoupling in solid-state NMR spectroscopy without dipolar truncation and influence from heteronuclear couplings using the triple oscillating field technique. When combined with rotor assisted dipolar refocusing, this new recoupling principle opens up new possibilities for accurate measurement of internuclear distances in extensive networks of coupled homonuclear spins as typically encountered for uniformly $^{13}$C (and $^{15}$N) labelled proteins. The easy readout of accurate internuclear distances via mapping in a grid of Fresnel curves indicates that our new dipolar recoupling concept may find widespread application in biological solid-state NMR spectroscopy. Furthermore, we anticipate that the triple oscillating field concept will pave the way for exciting new method developments for solid-state NMR structure determination.

\section{Acknowledgements}
This work was supported by the Danish National Research Foundation, the Danish Natural Science Research Foundation, and Danish Biotechnological Instrumentcentre (DABIC), the Office of Naval Research grant 38A-1077404, and the Air Force Office of Research Grants AFOSR FA9550-05-1-0443 and AFOSR FA9550-04-1-0427.

\end{document}